\documentstyle[aps,prb,twocolumn,epsf]{revtex}

\begin{document}

\title{ Effect of strain on the magnetic and transport
properties of the composite manganites,
La$_{2/3}$Ca$_{1/3}$MnO$_3$/yttria-stabilized zirconia}

\author{Wei Liu$^{1}$, Chinping Chen$^{1}$\cite{chen}, Xinfeng Wang$^{1}$, Jun
Zhao$^{1}$, Xiangyu Xu$^{1}$, Jun Xu$^{1,2}$, and Shousheng
Yan$^{1}$}

\address{
(1) Department of Physics and State Key Lab for Mesoscopic
Physics, Peking University, Beijing 100871, P.\,R.\,China\\(2)
Electron microscopy laboratory, Peking University, Beijing 100871,
P.\,R.\,China}
\maketitle

\begin{abstract}
The magnetic and transport properties have been investigated for
the composite polycrystalline manganites,
(1-x)$La_{2/3}Ca_{1/3}MnO_3$/(x)yttria-stabilized zirconia (
(1-x)LCMO/(x)YSZ ), at various YSZ fractions, x, ranging from 0 \%
to 15 \%. The ac magnetic susceptibility, $\chi(T)$, DC
magnetization, $M(T)$, temperature dependent resistivity,
$\rho(T)$ and thermoelectric power (TEP), $S(T)$, have been
measured. It was found, surprisingly, that a TEP peak showed up in
the magnetic transition region for the sample with the x even as
little as 0.75 \%. The magnetic transition temperature reaches the
minimum value as x increases from 0 \% to 4.5 \% and goes up as x
increases further. Several possible factors such as the effect of
strain, the finite size effect, and the effect of magnetic
tunnelling coupling, {\it etc.}, in affecting the above physical
properties of the composite manganites have been studied
carefully. The strain induced by the YSZ/LCMO boundary layer (BL)
was identified as the leading factor to account for the x
dependence of these properties. It demonstrates that the effect of
strain could be important in the bulk manganites as in the film
sample.

\end{abstract}


\section{Introduction}

The perovskite manganites, $A_{1-y}B_yMnO_3$, in which A is for
the rare earth trivalent cation and B for the alkaline divalent
one, exhibit complicated phases at various temperatures and hole
doping concentrations, y. Due to the important magnetic
application potentials and fundamental research interests,
tremendous activities in the physics community have focused in
this area during the past decade. With an appropriate doping
concentration y, an FM transition takes place at the Curie point,
$T_C$, with the decreasing temperature. It is accompanied with the
metal-insulator (MI) transition at $T_P$. A colossal
magneto-resistance (CMR) slao occurs around this temperature.
These properties can be explained with the double exchange (DE)
mechanism, \cite{Zener51,Anderson55,Goodenough55,deGennes60}.
\par
Recently, growing attentions have been focused on the effect of
strain arising from the interface or surface states of the CMR
thin film \cite{Dulli00,Bibes01,Bibes03}. The variation of the
magnetic transition temperature, $T_C$, was interpreted as
attributed to the strain. However, it is usually difficult to
separate the strain-induced effect from the finite size
confinement one with the thin films
\cite{Zhang01,Ziese02,Andres03}. On the other hand, the effect of
strain is usually overlooked in the bulk samples. We would like to
demonstrate that the effect of the BL strain with the
polycrystalline composite manganites, (1-x)LCMO/(x)YSZ, is a very
important factor for the variation of the physical properties such
as $T_C$, the temperature dependent resistivity, $\rho(T)$, and
the $S(T)$ behavior, {\it etc.}.

\section{Sample preparation}

A double-staged process was applied in preparing the LCMO/YSZ
samples \cite{Yuan02}. In the first stage, the LCMO nano-sized
powder was produced by the sol-gel method and then sintered at
1300$\ ^{o}C$ for 10 hours to form crystals of about 20 $\mu$m. In
the second stage, the thus-obtained LCMO powders were mixed with
the YSZ powders of about 2 $\mu$m for the heat treatment at 1350$\
^{o}C$ for another 10 hours. The X-ray diffraction (XRD) was
performed by a Philip x' pert diffractometer using the Cu
K$_{\alpha}$ line (1.54056 $\AA$). The XRD spectra are presented
in Fig.~\ref{XRD}. The YSZ phase was identified for the samples
with x exceeding 4.5 \%. The lattice constants of the LCMO phase
remain unchanged within 0.001 $\AA$ for all of the YSZ mixing
concentration, x. This indicates that, within the detection
sensitivity of XRD, the bulk LCMO composition was not modified
during the heat treatment due to any possible diffusion of ions
from the YSZ composition. We have prepared the samples of LCMO
annealed at 1400$\ ^{o}C$ without the mixing of YSZ as well. This
would demonstrate the widely investigated disorder effect
resulting from different annealing conditions.
\par
The morphology of the sample was investigated by a scanning
electron microscope (SEM) performed on the system of FEI STRATA
DB235 focus ion beam (FIB) electron microscope. It revealed that
at low mixing concentration, x $<$ 4.5 \%, the LCMO phase formed
large cluster surrounded by a thin layer of YSZ component at a
thickness of the order of 10 nm. The BL area increases while the
cluster size of the LCMO phase decreases with the increasing YSZ
fraction. The interconnecting paths between the adjacent LCMO
clusters separated by the YSZ layers would reduce accordingly. On
the other hand, at the mixing fraction, x, exceeding about 4.5 \%,
the YSZ phase forms cluster-like structure by itself. Thus, the BL
area decreases correspondingly. This observation is consistent
with the previous report\cite{Yuan02}. Within the detection
sensitivities of the XRD and SEM, no indication of the inter
diffusion between the LCMO and the YSZ phases exists. The two
phases, hence, form solid mixture with the existence of the BL
strain in between. Since the YSZ phase is insulating and
non-magnetic. the LCMO/YSZ composites are, therefore, appropriate
for the investigation of the magnetic and transport properties
under the BL strain.

\section{Experiment}

The dc magnetization, $M(T)$, and ac magnetic susceptibility,
$\chi(T) = \chi'+ i\chi''$, were performed using Quantum Design
PPMS and MPMS, respectively. The ac susceptibility measurement was
carried out with the excitation field of 10 Oe at 113 Hz under a
few Oe of dc background field. The applied field for the $M(T)$
measurement, including the field-cooled (FC) and zero-field-cooled
(ZFC), is 50 Oe, while the field applied during the cooling stage
before the FC measurement is 3000 Oe. The magnetic transition
temperatures, $T_C(dc)$ and $T_C(ac)$, are obtained using the
definition, $dM/dT$ and $d\chi'/dT$. These two transition
temperatures agree with each other well within a few Kelvin, as
plotted in Fig.~\ref{mag1}. Also plotted in the same figure are
the metal-insulator (MI) transition temperature, $T_P$, determined
by the $\rho(T)$ measurement, and the magnetic transition
temperature width, $\Delta T$, defined by the difference of
temperatures at which $d\chi'/dT$ = 0. The correlation of the
$T_C$ with the BL effect is apparent in the figure. The lowest
$T_C$ occurs at x = 4.5 \%, corresponding to the minimum BL
inferred from the SEM observation. In order to investigate the
effect of thermal treatment on $T_C$, the sample of x = 0 annealed
at 1400$\ ^{o}C$ was measured as well for the magnetic transition
temperature, $T_C$. It is 260 K using the same maximum slope
criteria described above. The depression of $T_C$ down to about
245 K with the sample annealed at 1300$\ ^{o}C$ accounts for about
6 \% effect. This is a well studied effect and is attributed to
the disorder associated with the polycrystalline grain size
distribution, \cite{Sanchez96,Gupta96}.

\par
The out-of-phase component, $\chi''(T)$, is shown in
Fig.~\ref{chi2}. Two characteristic dissipation features, the peak
at high temperature, $T_{DH}$, and the bump at low temperature,
$T_{DL}$, appear for each sample and are depicted in the inset of
Fig.~\ref{chi2} as a function of the YSZ fraction, x. $T_{DH}$ is
lower than $T_C$ by a few Kelvins. It is resulting from the energy
dissipation process associated with the critical spin fluctuation
near the FM phase transition. The x-dependence of $T_{DH}$ is
therefore similar to that of $T_C$. On the other hand, $T_{DL}$
appears around 70 K for all of the samples, including the one with
x = 0. This indicates that the LT bump is independent of the BL.
In fact, similar bumps in $\chi''(T)$ at $T < T_C$ have been
observed in many previous experiments, attributed to the magnetic
inhomogeneity \cite{Moreira98}. Hence, the x independence of
$T_{DL}$ is an indication that the characteristic crystalline
grain size associated with the LCMO phase stays unaltered with the
mixing of the YSZ component. The field-cooled (FC) and
zero-field-cooled (ZFC) dc magnetization measurements on the
chosen samples, x = 0 \%, 1.25 \%, 4.5 \% and 15 \%, were
performed to investigate the magnetic inhomogeneity.
Fig.~\ref{FC-ZFC} shows the normalized magnetization, M(T)/M$_m$,
versus the reduced temperature, T/$T_C$, where M$_m$ is the
maximum magnetization occurring near the freezing temperature at
which the FC and ZFC branches of the $M(T)$ curves diverge. There
is no appreciable difference for the x = 0 \% sample from the
other ones, indicating that the magnetic disorder revealed by this
measurement is ascribed to the polycrystalline grains, independent
of the BL. This is of the same origin causing the LT bump in the
$\chi''(T)$ measurements. Note that, there is no structure in the
$M(T)$ curve or the in-phase component, $\chi'(T)$ (not shown
here), at the temperature near the LT bump. It indicates that it
is not resulting from a magnetic phase transition.

\par
The $\rho(T)$ measurement was carried out from 80 K to 300 K in a
home-made insert-probe by a standard 4-probe dc techniques using
cold-pressed indium as the electrical contacts. The typical
contact resistances is on the order of a few $\Omega$ with the
applied current on the order of a few mA. The $\rho(T)$ curves are
published in Fig. 1 by Liu {\it et al} \cite{Liu03}. In the region
below $T_C$, the $\rho(T)$ behaviors are analyzed with the various
scattering processes of the electrons by the function $\rho(T) =
\rho_0 + AT^a + BT^b$, where a = 2 or 3/2, and b = 5 or 9/2. The
maximum fitting range in temperature for a stable result is from
the lowest temperature of the measurement to $T \sim$ 0.8 $T_C$.
The $AT^a$ term with a = 2 or 3/2 gives equally good fitting of
the experimental data. The coefficient of the fitting, A,
corresponding to a = 3/2 or 2 also exhibits identical x
dependence. It is difficult to distinguish which of the following
two processes is the more important one, the $T^2$ term for the
electron-electron scattering or the $T^{3/2}$ term for the
scattering of electrons by the disordered spin glass component
\cite{SpinGlass93}. The x dependence of this term is represented
by the coefficient $A$, calculated using a = 2, versus x and
plotted in Fig.~\ref{RT}. The variation with different samples is
within a factor of 4. Similarly, an equally good fitting is
obtained with the $BT^b$ term using b = 5 for the electron-phonon
scattering or b = 9/2 for the electron-magnon scattering within
the framework of DE mechanism \cite{Kubo72}. Since the
$B$-variation versus x is the same using b = 9/2 or 5, the ratio
of $B/A$ is plotted in the inset of Fig.~\ref{RT} using b = 5.
Also plotted in the same inset is the $\rho_0/A$ ratio, by the
open squares. The x dependence of $\rho_0$ associated with the
residual or disorder scattering process is only slightly higher
than that of the $AT^a$ term. From the above analysis for the
various scattering processes, the $BT^b$ term is affected most
profoundly by the presence of the LCMO/YSZ BL layer, indicating
that the BL-induced strain has a strong effect on the
electron-phonon coupling strength.
\par
The temperature dependent TEP, $S(T)$, was measured with the
series of LCMO/YSZ samples by the home-made insert-probe using the
dc differential technique. The electrical contacts were made by
the cold-pressed indium. The sample was installed across two
temperature platforms. One was regulated at temperature $T$, while
the other controlled to vary within $T$ + $\Delta T$. A continuous
voltage output, $\Delta V$, taken by Keithley 2010 multimeter was
recorded with the corresponding $\Delta T$, typically a few tenths
of a kelvin, changing slowly. The slope of the linear relation
between $\Delta V$ and $\Delta T$, with the correction of the
contribution from the Cu leads, would give the measured TEP of the
sample. Abrupt TEP jumps occur during the magnetic phase
transition for the x $>$ 0 \% samples, but not for the sample with
x = 0 \%, see Fig.~\ref{TEP}. This demonstrates clearly a strong
correlation of the jumps with the existence of the BL, and was
interpreted as due to the magnetic inhomogeneity induced by the BL
strain \cite{Liu03}. Similar TEP jump with the magneto-TEP effect
has been observed in the thin film CMR manganites also
\cite{Jaime96,Jaime99}. The substrate strain unavoidably caused
the magnetic inhomogeneity in the sample. Under the applied
magnetic field, the inhomogeneous magnetic component has been
reduced. The TEP jump was therefore suppressed to show the
magneto-TEP effect.

\section{Discussion}

The non-magnetic, insulating YSZ component intermixing in the LCMO
causes variations in the magnetic and transport properties of the
manganites. Most of the interesting features in the physical
properties under current investigations are strongly correlated
with the LCMO/YSZ BL. Several effects would be introduced on the
samples by the existence of the BL. However, only a leading one is
accounted for the observed x dependence. The strain induced by the
BL would be the major factor identified in the present work.
\par
Usually, the effect of strain on the physical properties of the
manganites, especially on the $T_C$ variations, is studied with
the films. However, for thickness under a few hundred nanometers,
the finite size confinement effect would become important to
superimpose on the effect of strain. With the substrate strain,
the interface magnetic inhomogeneity has been directly observed by
the techniques of NMR \cite{Bibes01} or X-ray photoemission
spectroscopy \cite{Dulli00}. Nonetheless, for these films, the
finite size effect seems to be the leading factors in the
depression of $T_C$, dominating the effect of strain under
discussion. Fig.~\ref{FiniteSize} displays the shift of $T_C$
versus film thickness, $d$, summarized from many of the previous
experiments for various thin films grown on different
substrates\cite{Bibes01,Andres03,Gupta96,Bibes02,Campillo01,Snyder96,Huhtinen02,Xiong96,Rao98}.
The results follows very well the law of finite size
confinement\cite{Fisher72,FiniteSize00},
$|T_C(\infty)-T_C(d)|/T_C(\infty) = (\xi_0/d)^\lambda$ with
$\lambda$ = 1 in consistent with the result from the mean field
theory\cite{Zhang01}, where $\xi_0 =$ 6 nm is the correlation
length at T = 0 K, $T_C(d)$ is the transition temperature for a
film of thickness, $d$, and $T_C(\infty)$ is that for the
corresponding bulk samples. Note that the dispersion of the data
points in Fig.~\ref{FiniteSize} indicates that the effect of the
substrate strain and the crystallinity of the films superimposed
on top of the confinement effect are non-negligible. This is
reasonable since these points are summarized from various
experiments performed by different laboratories. The above result
indicates that the finite size effect is the leading factor for
the suppression of $T_C$ with the thin film samples at a thickness
less than a few hundred nanometers, even with the obvious
coexistence of the substrate strain often suggested as the solely
factor.
\par
For an LCMO cluster enclosed by the YSZ component at the small YSZ
fraction, x $\leq$ 4.5 \%, $T_C$ drops dramatically with the
increasing x. In this region, the cluster size is on the order of
several tens of micrometers. This is simply too large for the
finite size confinement effect to occur according to the analysis
for the thin films. For the YSZ serving as a non-magnetic
separation between the LCMO phase, the reduction in the effective
magnetic coupling is unlikely the major factor for the x
dependence of the $T_C$ depression either. This effect would more
or less level off at a layer thickness of a few nm according to
the previous experiment\cite{Sirena03}. In the present work, the
non-magnetic YSZ layer is at least 10 nm in thickness, reaching
the region of saturation for such an effect. The effect of
intergranular magnetic tunnelling coupling, which is beyond the DE
mechanism, is unlikely the major factor either responsible for the
observed properties. In this case, the depression of $T_C$ is less
than 5 \% with $T_P$ lower than $T_C$ by a temperature depending
on the extent of the intergranular coupling strength.
\cite{Sanchez96,Gupta96,Mahesh96,Mahendrian96,Hwang96,Yuan03}. The
main features of the present results, see Fig.~\ref{mag1}, do not
fit the description above since $T_C$ is depressed by more than 20
\% with the varying x and $T_P$ follows it closely, see
Fig.~\ref{mag1}. Furthermore, the insulating YSZ layer with a
thickness of more than 10 nm is too thick for the electrical
current to tunnel through at LT to show metallic property.
\par
In the polycrystalline LCMO/YSZ composite system, the LCMO cluster
is larger than 10 $\mu$m. The ratio of the boundary strained layer
over the volume depends on the thickness of the strained layer. It
is possible for the spatial relaxation of an interfacial strain to
extend over a distance of 1 $\mu$m \cite{Soh02}, and results in a
non-negligible volume fraction of the boundary strained layer in
the bulk LCMO component. According to the previous reports, $T_C$
would be seriously depressed by the biaxial strain resulting from
the substrate lattice mismatch. Merely 1 \% of the biaxial strain
would cause an order of 10 \% variation in $T_C$\cite{Millis98b},
as demonstrated by the experiment of ultrasound
spectroscopy\cite{Darling98}. Yet, such a low level of strain
would go undetected by the usual experimental techniques such as
the XRD analysis. The magnetic anisotropy or inhomogeneity caused
by the strain would explain the depression of $T_C$, and the
corresponding broadening of the magnetic transition, $\Delta T$.
\par
In the analysis of $\rho(T)$ at $T < T_C$, the residual term,
$\rho_0$, and the $AT^a$ term exhibit much less structure
dependence than the $BT^b$ term. This is a strong evidence
supporting that the electron-phonon coupling strength is modified
by the existence of the BL. Since the Jahn-Teller (J-T) phonon
mode depends strongly on the biaxial strain of the lattice
\cite{Millis98b}, it is reasonable to infer that the x dependence
of the $BT^b$ term is caused by the biaxial strain, which affects
the magnetic transition, $T_C$ as well.
\par
The strong correlation of the TEP jump during the magnetic
transition with the presence of the BL is interpreted as of
magnetic origin\cite{Liu03}, and can be reasonably explained by
the magnetic inhomogeneity induced by the strain. At LT, $S(T)$
shows a typical metallic behavior with a small absolute value. As
the temperature increases toward the HT region, the fraction of
the PM component having the semiconducting property increases. For
the x = 0 sample, the change in the PM fraction relative to the FM
one is smooth, showing a smooth transition in $S(T)$. On the other
hand, the introduction of the BL with the x $>$ 0 samples would
cause an extra contribution from the inhomogeneous magnetism,
resulting in an abrupt deviation of $S(T)$ from the metallic
region. Interestingly, in the previou work on the series of
samples with constant valence, $Pr_{0.7}Ca_{0.3-x}Sr_xMnO_3$
\cite{Hejtmanek96}, the temperature-dependent TEP behavior has
been demonstrated to correlate strongly with the magnetic
transition. Especially, a TEP jump begins at the temperature near
the ferromagnetic-antiferromagnetic (AF) transition as shown in
Fig.4 by Hejtmanek {\it et al} \cite{Hejtmanek96}. According to
the present picture in explaining the TEP behavior, the occurrence
of the AF component within the FM matrix is responsible for the
abrupt jump of the TEP. A noteworthy point, however, is that the
cause of the inhomogeneous distribution of the magnetism for the
$Pr_{0.7}Ca_{0.3-x}Sr_xMnO_3$ samples is attributed to the FM-AF
transition, quite different from the existence of the BL-induced
strain in the presence work.

\section{Conclusion}

In conclusion, the YSZ component in the LCMO/YSZ composite
materials induces a large effect on the various physical
properties such as the variations of $T_C$ and $T_P$, the
broadening of magnetic transition, the pronounced TEP jump during
the magnetic transition, and the resistivity variation in the LT
FM phase, {\it etc.}. The BL-induced strain plays a crucial role
in the explanation of the observed properties. In this respect,
the effect of strain is not only important in the manganite film
already reported by some of the recent works, but also has a
profound effect in the bulk sample, as demonstrated by the present
work.

\section{Acknowledgement}
We are grateful to Prof. Songliu Yuan of the Department of
Physics, Huazhong University of Science and Technology, Wuhan,
China, for providing us with the samples and for the helpful
discussions. One of the authors, C.P. Chen, would also like to
appreciate Prof. Tong-han Lin of the Department of Physics, Peking
University, Beijing, China, for some of the points raised and the
fruitful discussions.

\begin{figure}
\caption{XRD spectra for the series of samples, (1-x)LCMO/(x)YSZ.
The XRD was performed by a Philip x' pert diffractometer using the
Cu K$_{\alpha}$ line (1.54056 $\AA$). At x $\leq$ 4.5 \%, only the
LCMO phase is observed. The YSZ phase shows up at x $>$ 4.5 \%. }
\label{XRD}
\end{figure}

\begin{figure}
\caption{x dependence of Curie temperatures, $T_C(dc)$ by dc
magnetization and $T_C(ac)$ by ac susceptibility, metal-insulator
transition temperature, $T_P$, and magnetic transition width in
temperature, $\Delta T$.} \label{mag1}
\end{figure}

\begin{figure}
\caption{Out of phase component of magnetic susceptibility,
$\chi''$, versus temperature, T, for samples with various YSZ
concentration, x. The inset shows the x dependence of the two
dissipation characteristic temperatures, $T_{DH}$, corresponding
to the critical fluctuation around the ferromagnetic transition
and, $T_{DL}$, associated with the disorder spin state on the
polycrystalline grain surface. } \label{chi2}
\end{figure}

\begin{figure}
\caption{FC-ZFC dc magnetization. M$_m$ is the maximum
magnetization occurring near the freezing temperature. The solid
symbols are for the FC results, and the open ones are for the ZFC
branches. } \label{FC-ZFC}
\end{figure}

\begin{figure}
\caption{x dependence of the scattering processes of electrical
transport at $T<T_C$. The resistivity is fitted by the equation,
$\rho(T)=\rho_0+AT^2+BT^5$ to obtain $\rho_0$, $A$, and $B$. The
fitting range in temperature is $T < 0.8  T_C$. The solid circles
in the inset represent the ratio of $B/A$, which show variation
with more than an order of magnitude.} \label{RT}
\end{figure}

\begin{figure}
\caption{Temperature dependent TEP of the samples with various YSZ
fraction, x. The solid circle is for x = 0\%, while the open ones
for samples with x $>$ 0. } \label{TEP}
\end{figure}

\begin{figure}
\caption{Finite size confinement effect on the shift of magnetic
transition temperature according to the equation,
$|T_C(\infty)-T_C(d)|/T_C(\infty) = (\xi_0/d)^\lambda$, for
various thin manganite films on different substrates. The solid
line is calculated using the above equation with $\lambda$ = 1 and
$\xi_0$ = 6 nm. The manganite films include,
$La_{2/3}Ca_{1/3}MnO_3$ (LCMO), $La_{2/3}Sr_{1/3}MnO_3$ (LSMO),
$Pr_{2/3}Sr_{1/3}MnO_3$ (PSMO), and $La_{0.8}Ca_{0.2}MnO_3$, while
the substrates are $SrTiO_3$ (STO), $LaAlO_3$ (LAO), and $NdGaO_3$
(NGO). These results are summarized from the experiments,
(a)\cite{Bibes02}, (b)\cite{Bibes01}, (c)\cite{Andres03},
(d)\cite{Gupta96}, (e)\cite{Campillo01}, (f)\cite{Snyder96},
(g)\cite{Huhtinen02}, (h)\cite{Xiong96}, (i)\cite{Rao98}.  }
\label{FiniteSize}
\end{figure}

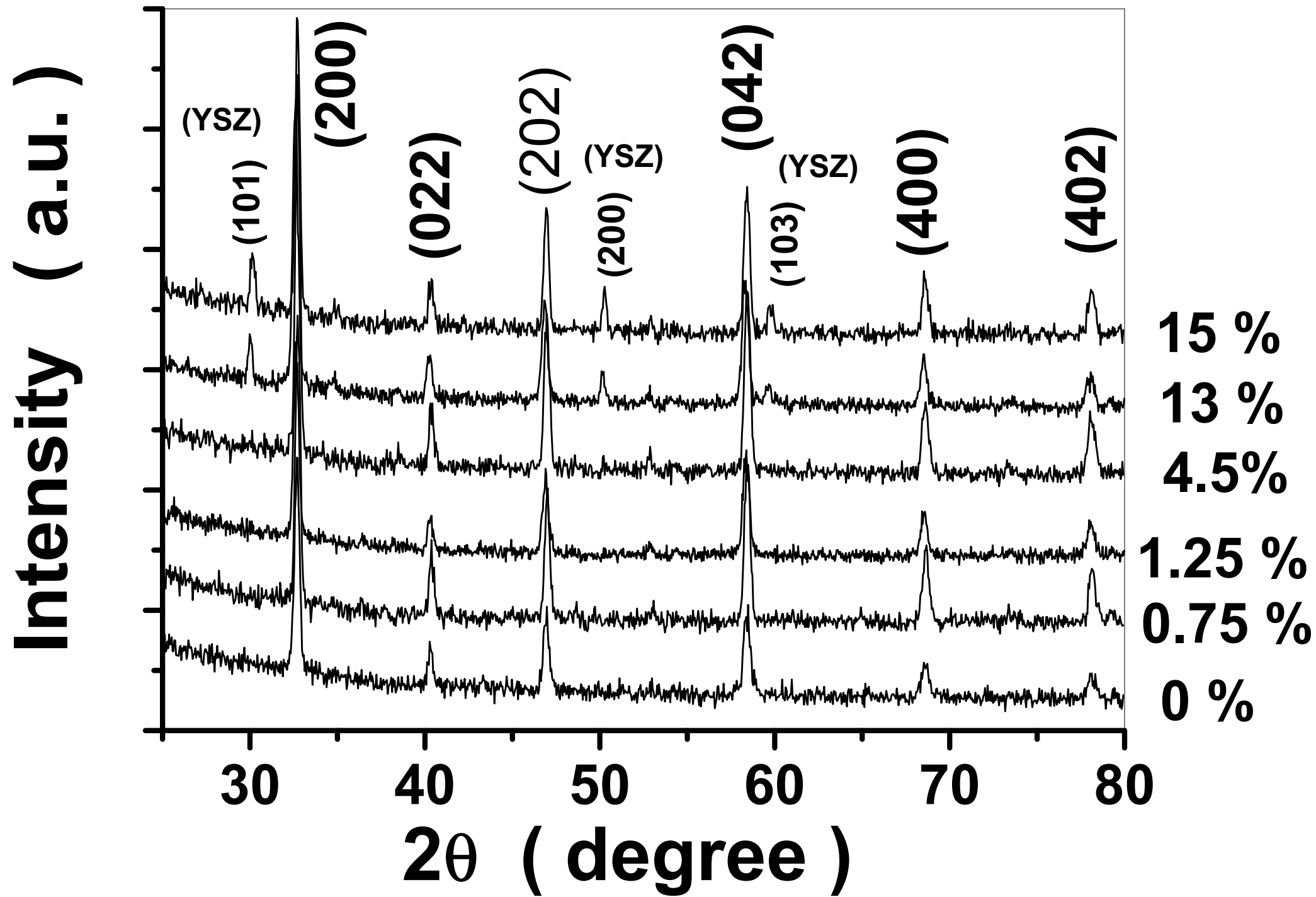
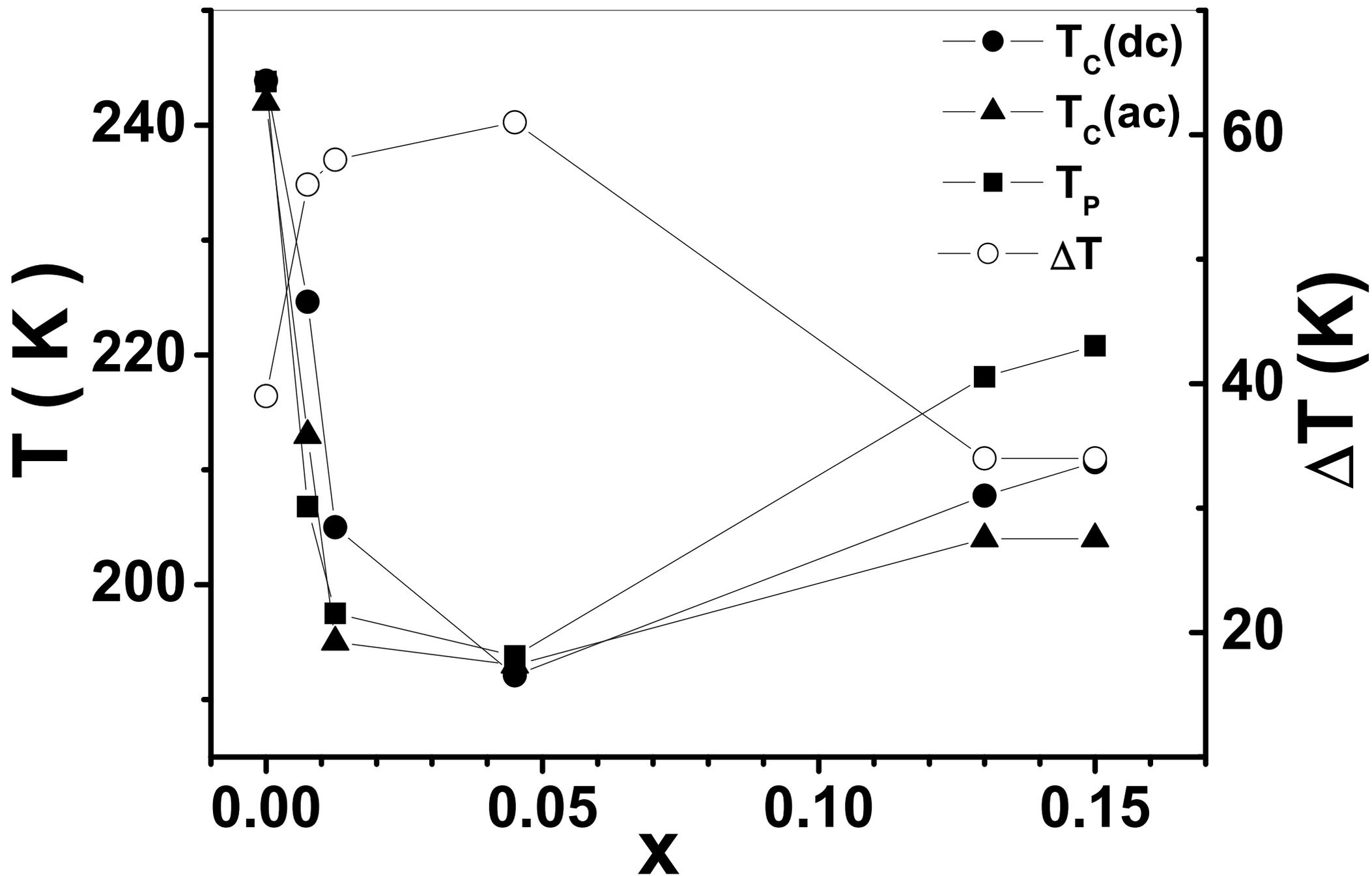
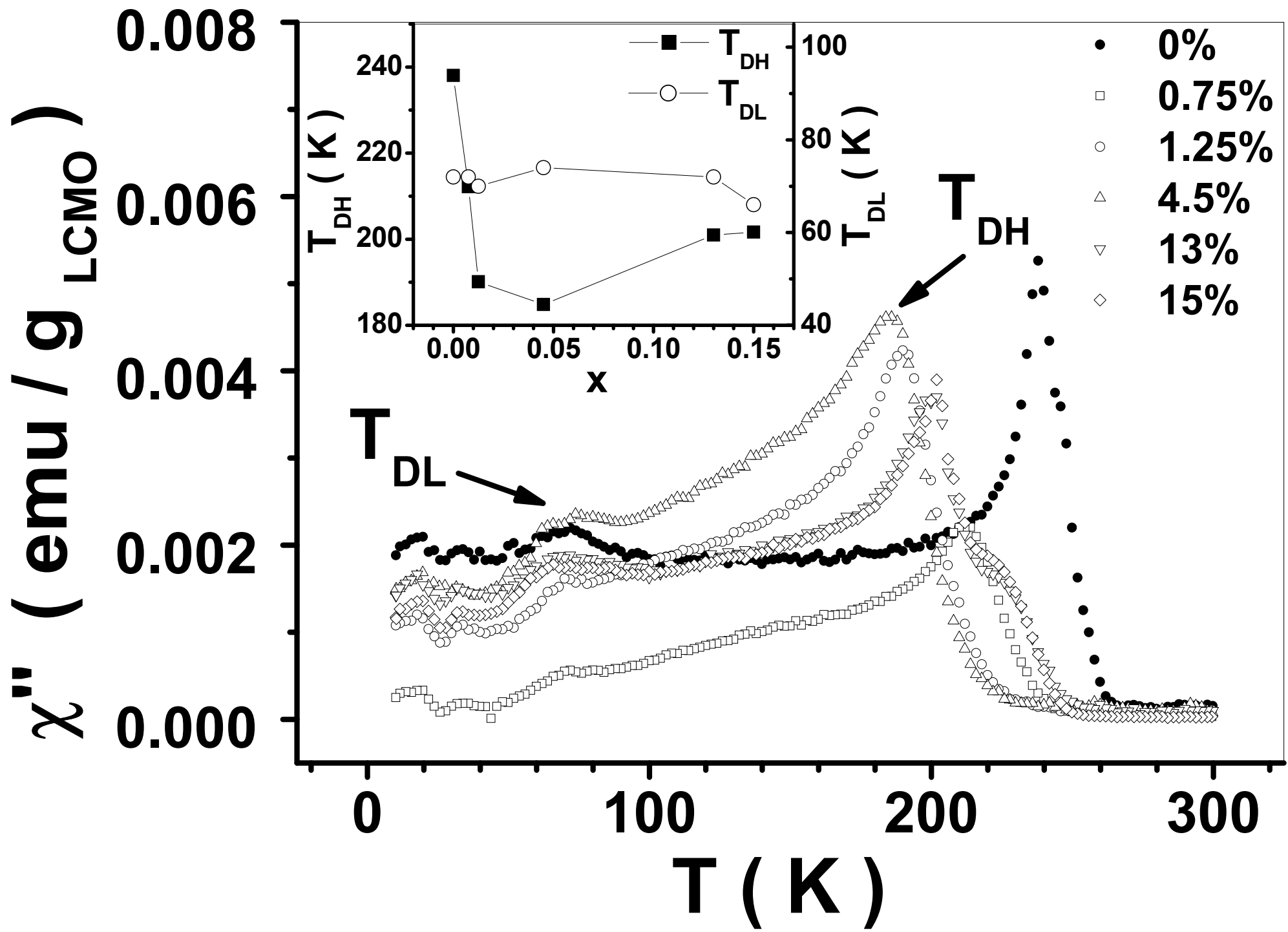
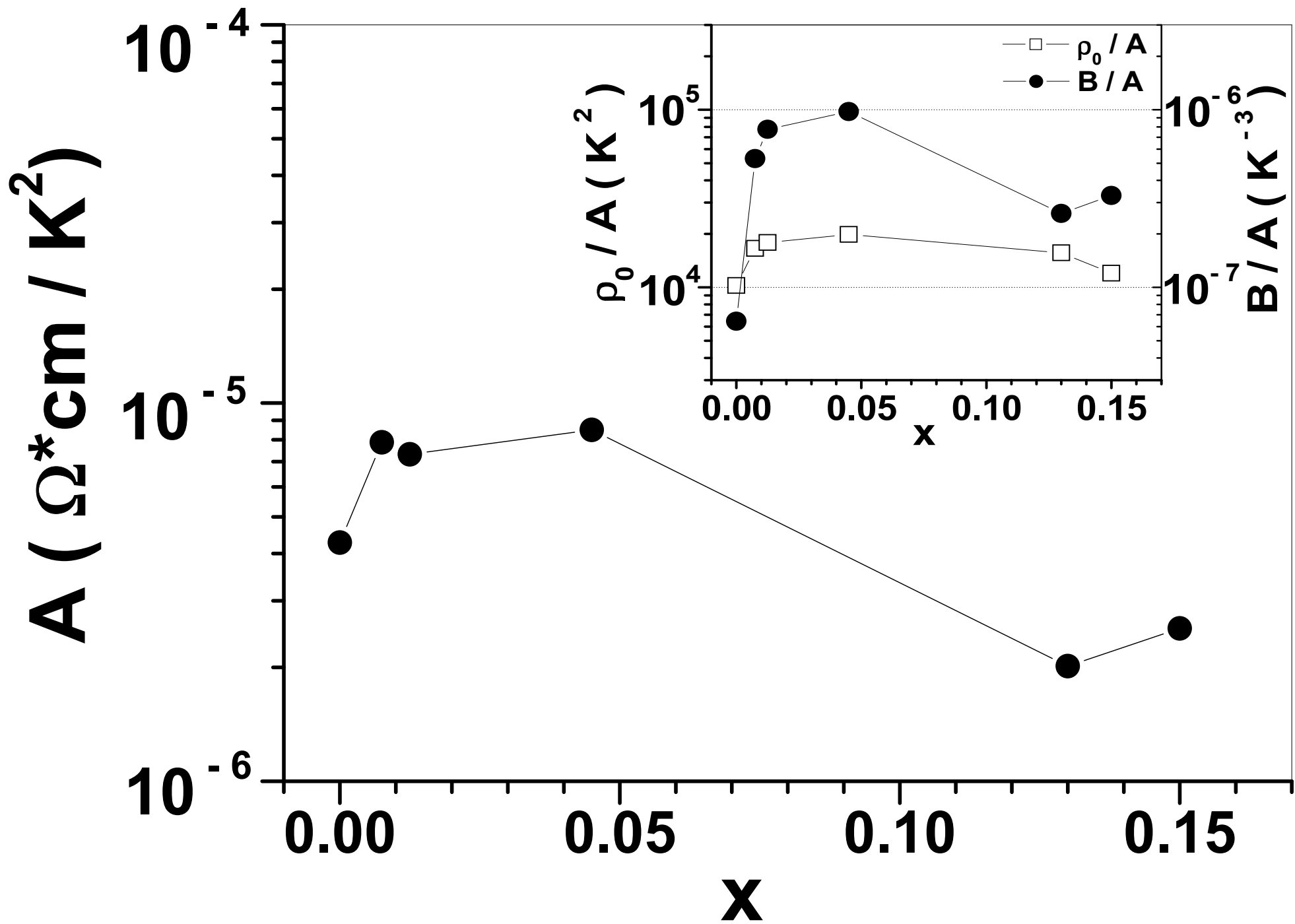
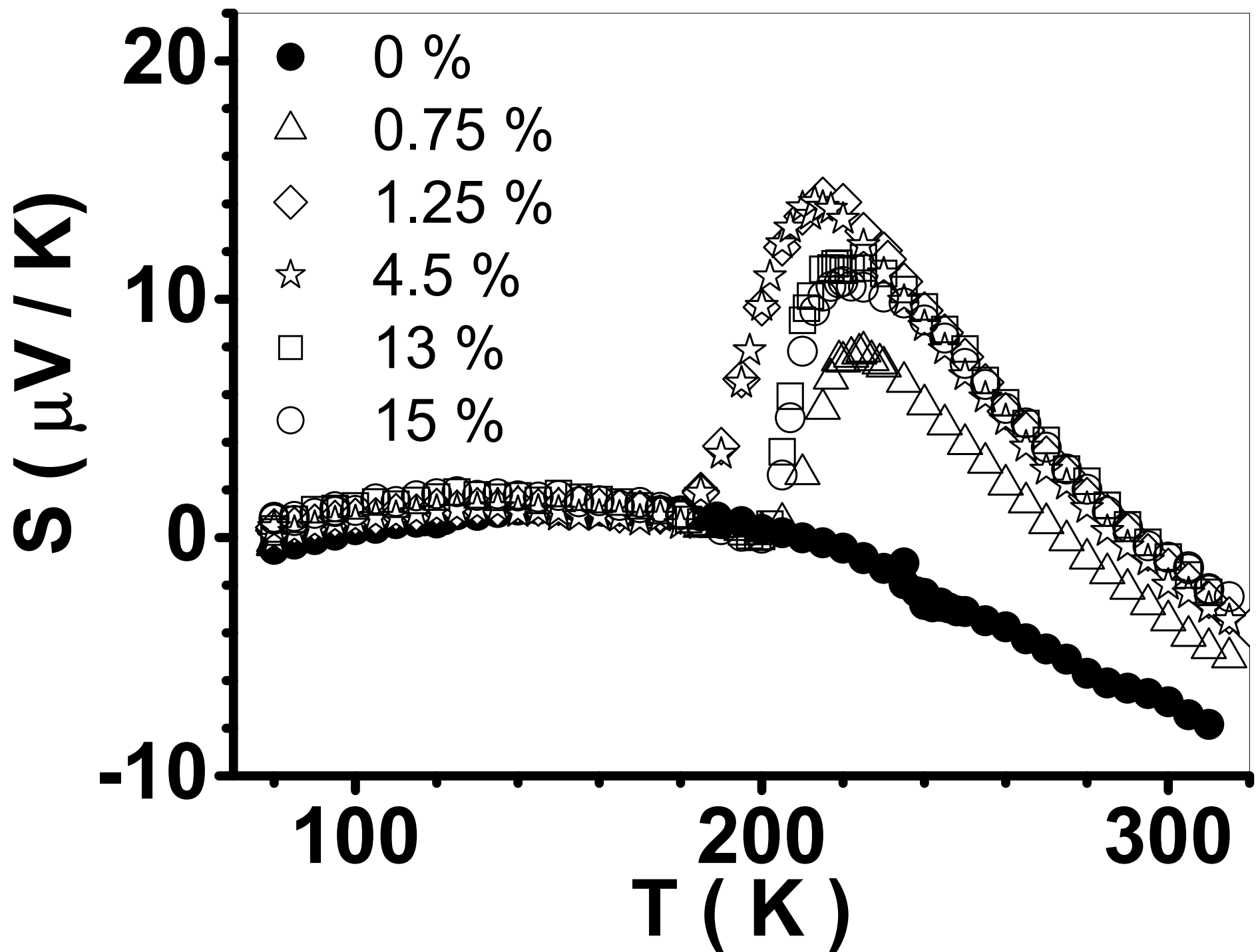
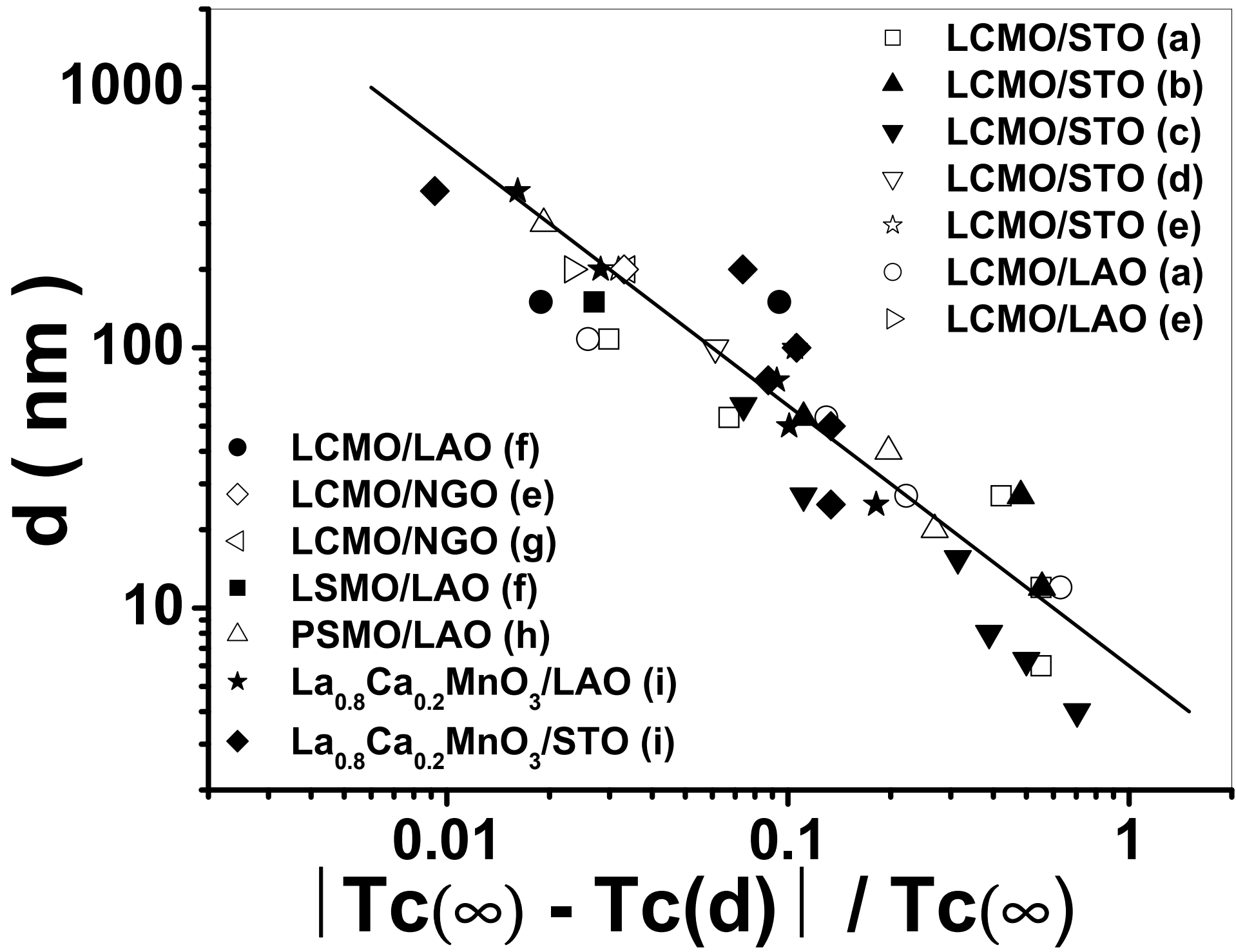
\end{document}